\documentclass{svjour2}                    
\smartqed  
\usepackage{graphicx}
\usepackage{amssymb}
\usepackage{amsmath}
\usepackage{pst-all}
\usepackage{epsfig}
\usepackage{rotating}
 
\newcommand{\be}{\begin{equation}}
\newcommand{\ee}{\end{equation}}
\newcommand{\bea}{\begin{eqnarray}}
\newcommand{\eea}{\end{eqnarray}}
\newcommand{\lb}{\label}
\newcommand{\bdm}{\begin{displaymath}}
\newcommand{\edm}{\end{displaymath}}
\newcommand{\D}{{\rm d}}
\newcommand{\E}{{\rm e}}
\newcommand{\I}{{\rm i}}

 \journalname{Gen Relativ Gravit}
\begin{document}

\title{Quantum Geometrodynamics:\\ whence, whither?\thanks{Dedicated to the
memory of John Archibald Wheeler.}
}

\titlerunning{Quantum Geometrodynamics}        

\author{Claus Kiefer}

\institute{Claus Kiefer 
	\at Institut f\"ur Theoretische Physik, 
Universit\"at zu K\"oln, Z\"ulpicher Stra\ss e 77, 50937 K\"oln, Germany\\
	\email{kiefer@thp.uni-koeln.de}
}

\date{Received: date / Accepted: date}

\maketitle

\begin{abstract}

Quantum geometrodynamics is canonical quantum gravity with the
three-metric as the configuration variable. Its central equation is
the Wheeler--DeWitt equation. Here I give an overview of
the status of this approach. The issues discussed
include the 
problem of time, the relation to the covariant theory, the
semiclassical approximation as well as applications to black holes and
cosmology. I conclude that quantum
geometrodynamics is still a viable approach and provides insights into
both the conceptual and technical aspects of quantum gravity. 


\end{abstract}

\vskip 1cm

\begin{quote}
These considerations reveal that the concepts of spacetime and time
itself are not primary but secondary ideas in the structure of
physical theory. These concepts are valid in the classical
approximation. However, they have neither meaning nor application
under circumstances when quantum-geometrodynamical effects become
important. \ldots There is no spacetime, there is no time, there is no
before, there is no after. The question what happens ``next'' is
without meaning. 
\end{quote}
(John A. Wheeler, {\em Battelle Rencontres} 1968)

\newpage


\section{Introduction}

The quantization of the gravitational field is still among the most
important open problems in theoretical physics. Despite many attempts,
a final theory, which has to be both mathematically consistent and
experimentally tested, remains elusive. John Wheeler once wrote: ``No
question about quantum gravity is more difficult than the question,
`What is the question?''' \cite{Wheeler}. One of the questions is, of
course, which approach to quantum gravity one is motivated to pursue.

My contribution here is devoted to one particular approach -- quantum
geometrodynamics. Being one of the oldest, it is still an active field
of research. Quantum geometrodynamics is one version of canonical
quantum gravity, to which also loop quantum gravity belongs. All
canonical theories contain as their central equations {\em
  constraint equations}, that is, quantum versions of classical
constraints between the generalized positions and momenta of the
theory.
In the case of gravity, these are the
Hamiltonian and diffeomorphism constraints augmented, in the case of
the loop approach, by the Gauss constraints. But the
various canonical approaches are distinguished by
their choice of canonical variables: three-metric and extrinsic
curvature in geometrodynamics, holonomies and fluxes in the loop
version. The non-trivial relationship between the various canonical
variables leads to different, most probably inequivalent, quantum
theories with different mathematical structures. Only the experiment
can decide, at the end, which of them is the correct one, if any. 

All the canonical theories are approaches which focus on the direct
quantization of Einstein's theory of general relativity. They thus do
not necessarily entail a unification of gravity with the other
interactions. Alternative approaches to a quantum theory of relativity
are the covariant ones to which standard perturbation theory 
and path-integral quantization belong. Fundamentally different in
spirit is string theory whose major aim {\em is} a unification of all
interactions within one quantum framework. Quantum gravity as such
emerges there only in an appropriate limit in which the various
interactions becomes distinguishable. An introduction to all major
approaches can be found in my monograph \cite{OUP}. The reader can
also find there a more complete list of references.

The purpose of this contribution is to provide a concise and critical
review of the status of quantum geometrodynamics, its successes and
shortcomings. I shall start in Section~2 with a brief introduction to
the formalism 
of canonical gravity at both the classical and quantum level. I
discuss in particular the problem of time and the relation of
geometrodynamics to the covariant approaches. A brief historical
overview is also
included. Section~3 focuses on one of the successes: the relation of
quantum geometrodynamics to quantum theory on a fixed background. This
concerns in particular the recovery of the (functional) Schr\"odinger
equation and its quantum gravitational corrections. Sections 4 and 5
then give a brief overview of the main applications: quantum black
holes and quantum cosmology. I shall end with some conclusions and an
outlook.


\section{What is quantum geometrodynamics?}

\subsection{The 3+1-decomposition}

The usual starting point for developing the canonical formalism is
the foliation of spacetime into three-dimensional spacelike
hypersurfaces. A 
prerequisite for this is the global hyperbolicity of the
spacetime. Figure~1 shows schematically two infinitesimally
neighboured hypersurfaces. The vector $\dot{X}^{\mu}\D t$,
where
\be
\dot{X}^{\nu}\equiv t^{\nu}=Nn^{\nu}+N^aX^{\nu}_{,a}\ ,
\ee
denotes the connection between points with the same spatial
coordinates $x^a$. This connection can be decomposed into a normal and
a tangential part. The amount of the normal separation is specified by
the lapse function $N$ (with $n^{\mu}$ denoting a unit normal vector);
the tangential separation is quantified by the components $N^a$ of the
shift vector. 
 \begin{figure}[h]
  \begin{center} 
  \includegraphics[width=0.7\textwidth]{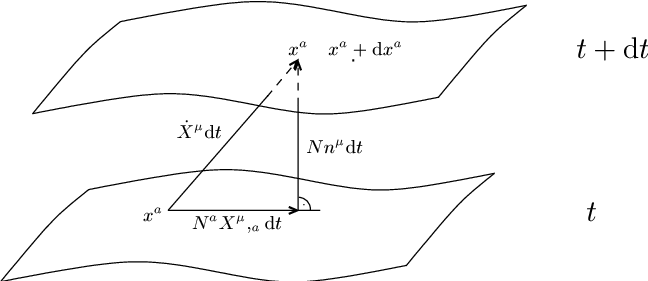} 
  \caption{Two successive spacelike hypersurfaces in the 3+1-decomposition.}
  \end{center}
\end{figure}
The four-dimensional line element between a point with coordinates
$x^a$ on the lower hypersurface to a point with coordinates $x^a+\D
x^a$ on the upper hypersurface can then be decomposed as follows:
\bea
\D s^2 &=& g_{\mu\nu}\D x^{\mu}\D x^{\nu}
=-N^2\D t^2+h_{ab}(\D x^a+N^a\D t)(\D x^b+N^b\D t)\nonumber\\
&=& (h_{ab}N^aN^b-N^2)\D t^2+ 2h_{ab}N^a\D
  x^b\D t+h_{ab} 
\D x^a\D x^b\ ,
\eea
where $h_{ab}$ denotes the components of the three-dimensional metric,
in brief: the three-metric. In the canonical formalism, the three-metric
will play the role of the configuration variable. 
To quote again John Wheeler: ``The formalism of quantum gravity, in
its best developed form, makes three-geometry a central concept''
\cite{Wheeler}. Instead of
considering a three-metric on each hypersurface, we can imagine a
given three-manifold $\Sigma$ and a $t$-dependent three-metric on
it. In fact, the canonical formalism depends on the chosen manifold
$\Sigma$; there is one canonical theory for each $\Sigma$.

This leads to a more fundamental viewpoint, cf. \cite{GK}. We can
assume that in the beginning {\em only} $\Sigma$ is given, not a
spacetime. Only {\em after} solving the dynamical equations are we
able to construct spacetime and {\em interpret} the time dependence of
the metric $h_{ab}$ on $\Sigma$ as being brought about by `wafting'
$\Sigma$ through a four-manifold via a one-parameter family of
embeddings. 

The classical equations are six evolution equations for
the $h_{ab}$ and their momenta $p^{ab}$ as well as four {\em
  constraints} for them. The momenta $p^{ab}$ are linear combinations
of the extrinsic curvature of the three-dimensional space. The six
evolution equations and four constraints are the canonical version of
the ten Einstein field equations. Only after the classical equations
have been solved, can one interpret spacetime as a `trajectory of
spaces'. 

In the quantum theory, the trajectories will disappear as in ordinary
quantum mechanics. There will thus be no spacetime at the most
fundamental level;
only the constraints for the three-dimensional space will remain. But before
discussing them in the quantum theory, we shall have a brief look at their
classical version. 


\subsection{Constraints}

As mentioned in the last subsection, 
Einstein's equations can be written as a dynamical system of evolution
equations together with constraints. The constraints are, at each
space point, the {\em Hamiltonian constraint} $H\approx 0$ and
the three {\em momentum} or {\em diffeomorphism} constraints
$D^a\approx 0$, where $a=1,2,3$. The sign $\approx$ denotes here
the weak equality of Dirac, according to which the constraints can
be used only after the evaluation of Poisson brackets. The explicit
form of the constraints reads,
\begin{alignat}{2}
& H[h_{ab},p^{cd}] &&\,=\,2\kappa\,G_{ab\,cd}p^{ab}p^{cd}
\,-\,(2\kappa)^{-1}\sqrt{h}({}^{\scriptscriptstyle (3)}\!R-2\Lambda)\,
+\,\sqrt{h}\rho\approx 0\,, \lb{H}\\
& D^a[h_{ab},p^{cd}] &&\,=-2\nabla_bp^{ab}\,+\,\sqrt{h}j^a\approx 0\,,\lb{Da}
\end{alignat}
where $h$ is the determinant of the three-metric,
${}^{\scriptscriptstyle (3)}\!R$ the three-dimensional Ricci scalar,
$\Lambda$ the cosmological constant,
$\rho$ ($j^a$) denotes the energy density (current) of the
non-gravitational fields and
\bdm
\kappa=8\pi G/c^4\ .
\edm 
The coefficients $G_{ab\,cd}$ denote the
``DeWitt metric'' and are explicitly given by
\be
G_{ab\,cd}=\tfrac{1}{2\sqrt{h}}(h_{ac}h_{bd}+h_{ad}h_{bc}-h_{ab}h_{cd})\ .
\ee
The configuration space on which the constraints are defined is the
space of all three-metrics and is called Riem $\Sigma$. The
interpretation of the diffeomorphism constraints \eqref{Da} is
straightforward: they generate spatial coordinate transformations on
$\Sigma$. What really counts is therefore the space of all three-{\em
  geometries}, which is obtained from Riem $\Sigma$ after dividing out
the diffeomorphisms. This space of all three-geometries has been
baptized {\em superspace} by John Wheeler (it has nothing to do with
supersymmetry) and is sometimes 
considered to be the real configuration space of canonical gravity.
Its mathematical structure is highly non-trivial, see, for example,
\cite{OUP,GK} and the references therein. 

If the three-dimensional space $\Sigma$ is compact without boundary,
the full Hamiltonian is a sum of the above constraints. In the
asymptotically flat case, it contains in addition boundary terms
coming from the Poincar\'{e} charges at infinity, which include the
ADM energy \cite{ADM}. 

The Hamiltonian constraint can be mathematically interpreted as the
generator of normal hypersurface deformations, that is, of deformations
normal to the spacelike hypersurfaces in the canonical
formalism. Together with \eqref{Da}, it obeys the Poisson constraint
algebra of all hypersurface deformations (normal and tangential)
\cite{OUP}. This symmetry is not equivalent to the four-dimensional
symmetry of spacetime diffeomorphisms; however, the Hamiltonian
formalism together with the hypersurface deformations is equivalent to
the Lagrangian formalism with the spacetime diffeomorphisms. The
constraint algebra closes, that is, the Poisson bracket between two
constraints is proportional to a linear combination of the
constraints. It is not a Lie algebra, though, because the Poisson
bracket between two Hamiltonians \eqref{H} contains
on the right-hand side explicit functions of the canonical variables.

There exists a subtle and intriguing connection between 
the constraints and the dynamical evolution \cite{Kuchar,GK}.
Firstly, the constraints are preserved in time if and only if the
energy--momentum tensor of matter has vanishing covariant
divergence. This has an analogon in electrodynamics: the Gauss
constraint is preserved in time if and only if the electric charge is
conserved. Secondly, Einstein's equations are the unique propagation
law consistent with the constraint: if the constraints hold on every
hypersurface, Einstein's equations hold on spacetime; conversely, if
the constraints are valid on a particular hypersurface and if
Einstein's equations hold on spacetime, the constraints hold on every
hypersurface. This possesses, again, an analogon in electrodynamics:
Maxwell's equations are the unique propagation law consistent with the
Gauss constraint. In a sense, the dynamical equations in general
relativity follow entirely
from the ``laws of the instant'', that is, from the constraints
\cite{Kuchar}. 


\subsection{Problem of time I}

The fact that the laws of the instant suffice gives rise to the
classical facet of the {\em problem of time}, cf. \cite{Barbour94}.  
Let us restrict attention, for
simplicity, to a compact three-space $\Sigma$. The total Hamiltonian
is then a combination of the constraints only: the whole evolution
is generated by the constraints. This shows again that the
dynamical laws follow entirely from the constraints. No external time
parameter exists, and all physical time variables, if needed, 
must be constructed
from within the system, that is, as a functional of the canonical
variables. (Such physical time variables may come into play upon
solving the constraints.)
A priori, there is no preferred choice of such an intrinsic
time parameter. Still, in the classical theory a spacetime can be
constructed after solving the field equations and can thus be
described by a classical time function. This is no longer possible in
the quantum theory where the spacetime itself (the ``trajectory of spaces'')
vanishes, giving rise to a more fundamental problem of time, see
below.  

The problem of time is connected with the problem of observables. The
status of the latter is a subject of debate. The concept of
observables was introduced by Peter Bergmann into the field of
constrained dynamics to denote variables which have vanishing Poisson
brackets with all of the constraints. Since constraints are believed
to generate redundant (``gauge'') transformations, these variables would
be invariant under such transformations and would thus be candidates
for physical variables. In fact, Bergmann coined the name observables
in the hope that after quantization they would play the role of what
is called observables in quantum theory. 

This notion of observables may indeed be the appropriate one for gauge
theories. It has, however, been disputed whether it is also the
appropriate one for the situation encountered here \cite{Kuchar,BF}.
A quantity having vanishing Poisson brackets with both the Hamiltonian
and the diffeomorphism constraints (i.e. a quantity
``commuting'' with them) is a constant of motion because, at least in
the spatially compact case, the full Hamiltonian is the sum of these
constraints. This is another aspect of the problem of time -- no
time, no motion. 

In order to avoid such a far-reaching conclusion,
Kucha\v{r} has introduced the alternative concept of a
{\em perennial} for a quantity that commutes with all constraints,
that is, with both \eqref{H} and \eqref{Da}, and has instead reserved
the notion {\em observable} for a quantity that commutes only with the
diffeomorphism constraints \eqref{Da} \cite{Kuchar}. 

This makes sense. As Barbour and Foster have convincingly argued, it
is misleading to think of the Hamiltonian constraint \eqref{H} as a
generator of pure gauge transformations \cite{BF}.
To support this claim they have
focused on a particle model with a Hamiltonian constraint, where
this constraint only generates reparametrizations of the curve
parameter. They show that the presence of this constraint has to do
with the fact that the initial condition for a geodesic in
configuration space is a point and a direction at that point, not the
absolute value of a velocity, and that the Hamiltonian does generate
physical change. Extrapolating this insight to the gravitational
situation, one would conclude that physical quantities are only
required to commute with the diffeomorphism constraints \eqref{Da},
that is, that they do not need to be perennials. The Hamiltonian constraint
yields a transformation from one configuration to a different one.


\subsection{Quantization}

Within the canonical formalism one can employ two approaches towards
quantization. In the first one, 
one tries to solve the constraints \eqref{H} and \eqref{Da} at the
classical level in order to arrive at a formulation with
unconstrained, ``physical'', variables only. This is called reduced
quantization. In practice, this approach is hardly feasible; it is
even in quantum electrodynamics impossible to work with a reduced
formulation -- only in the non-interacting case can one identify the
free transversal fields as the unconstrained variables. 

One thus usually follows the second path, which is Dirac
quantization \cite{Dirac50}. 
In general, one would not expect this approach to be equivalent with
reduced quantization, cf. \cite{Gotay}. However, using path-integral
methods (cf. Section~2.6) one can show that at least in the one-loop
(linear in $\hbar$-) approximation, reduced and Dirac quantization are
equivalent if a particular factor ordering for the operators is chosen
\cite{b1,b2,unitary}. 

Let us focus on Dirac quantization.
Poisson brackets of the canonical
variables are translated into commutators, and the classical
constraints are translated into restrictions on physically allowed
wave functions. In Dirac's words (\cite{Dirac50}, p.~145): 
\begin{quote}
Weak equations between the classical variables correspond to linear
conditions on the vectors $\psi$, according to the formula
\bdm
X(q,p)=0 \quad {\rm corresponds\ to} \quad X\psi=0.
\edm
\end{quote} 
(The weak equality sign $\approx$ for the constraints was introduced
later \cite{Dirac58}.)
In our case it is the three-metric and its canonical momentum which
in the quantum theory obey the canonical commutation relation. In the
Schr\"odinger representation, the components of the momentum are
substituted by $\hbar/{\rm i}$ times the functional derivative with
respect to the metric,
\be
\lb{momentum}
\hat{p}^{ab}\longrightarrow \frac{\hbar}{{\rm i}}\frac{\delta}{\delta
h_{ab}}\ .
\ee
This is a formal heuristic prescription only, since one cannot expect
the momentum to be represented by a self-adjoint operator, the reason
being its non-commutation with the constraints. 

In fact, the rule \eqref{momentum} does not implement one important
property of the three-metric: the positivity property that demands
${\rm det}\ h_{ab}>0$. It has thus been suggested to replace
\eqref{momentum} by a modified prescription, leading to a variant of
the canonical approach known as affine quantization
\cite{Klauder}. The question as to which prescription is correct
has to do with the problem of factor ordering.

With these formal rules,
the classical Hamiltonian constraint \eqref{H} becomes the
quantum Hamiltonian constraint, also known as the {\em Wheeler--DeWitt
  equation} \cite{DeWitt,battelle}, 
\be
\lb{WDW}
\hat{H}\Psi\equiv
\left(-2\kappa\hbar^2G_{abcd}\frac{\delta^2}{\delta h_{ab}\delta h_{cd}}
-(2\kappa)^{-1}\,\sqrt{h}\bigl(\,{}^{(3)}\!R-2\Lambda\bigr)
\ +\,\sqrt{h}\hat{\rho}\right)\Psi=0\ .
\ee
Similarly, the diffeomorphism constraints \eqref{Da} are translated
into their quantum version, 
\be
\lb{qdiffeo}
\hat{D}^a\Psi \equiv -2\nabla_b\frac{\hbar}{\rm i}
\frac{\delta\Psi}{\delta h_{ab}}\,+\,\sqrt{h}\hat{j}^a\Psi =0\ .
\ee
In these equations, a ``naive'' factor ordering has been chosen in
the sense that all momenta are written to the right of the
metric-dependent terms. 
The argument of the quantum geometrodynamical wave functional $\Psi$
is the three-metric $h_{ab}$ together with the non-gravitational
degrees of freedom defined on $\Sigma$ (in the simplest situations
taken to be a
scalar field). It is easy to see that \eqref{qdiffeo} guarantees that
the wave functional is independent under a spatial coordinate
transformation which is connected with the identity. (It can acquire a
phase under a so-called large diffeomorphism.) A similar feature is
the quantized Gauss constraint in electrodynamics and Yang--Mills
theories, which guarantees the invariance of the wave functional under
infinitesimal gauge transformations. 

As they are written down, the
equations \eqref{WDW} and \eqref{qdiffeo} are of a formal nature only,
that is, they require a precise mathematical formulation. Such a
formulation is not 
yet available, except within a one-loop approximation scheme,
cf. Section~2.6. Using a different set of canonical variables, one
arrives at alternatives, most likely inequivalent, versions of
canonical quantum gravity. One of them is
loop quantum gravity, which has its own advantages and shortcomings,
see \cite{Rovelli} as well as other contributions to this volume. 

The mathematical
problems of quantum geometrodynamics have to do with factor
ordering, regularization, 
and Dirac consistency, which are themselves intertwined problems. The latter
refers to the quantum version of the classical constraint algebra, for
which it is not clear that it closes on the constraints in the way the
classical algebra does; the algebra may contain additional
`anomalous' terms.
 If it does not close, the equations \eqref{WDW} and
\eqref{qdiffeo} will not be consistent because the anomaly would yield
a non-vanishing term. This is what happens, in fact, for the quantum
Virasoro algebra in string theory, where it is of the utmost
importance. It is not at all obvious that such an anomaly is absent
for geometrodynamics. This question can, of course, only be
consistently addressed after the constraints have been regularized. 
It must be emphasized that many of these problems are not peculiar to
quantum geometrodynamics, but occur in other approaches as well, in
which general relativity is directly being quantized. 

The main purpose of the equations \eqref{WDW} and \eqref{qdiffeo} is
then twofold: on the one hand, it can give intuitive insight by formal
manipulations of the equations. On the other hand, they may be
truncated into well-defined equations in the context of particular
models, notably in quantum cosmology. 
We shall encounter applications of both kinds below.
In most of the formal
applications as well as in the concrete models, the subtle features
connected with the choice of factor ordering and possible anomalies is
less relevant. Situations where they {\em are} definitely of relevance
include discussions of the singularity avoidance in quantum gravity.


\subsection{Problem of time II}

In the quantum theory, the problem of time becomes more pressing. Not
only the external time, but also spacetime as such has diappeared!
This conclusion is unvoidable as long as one sticks to the usual
quantum formalism (as we do here). In quantum mechanics, particle
trajectories are absent. In quantum gravity, spacetime is the
entity that is analogous to a particle trajectory; consequently, it is
absent at the most fundamental level. In the canonical formalism
discussed so far, space (in the form of the three-dimensional
manifold) still exists. It may acquire a discrete structure (as
seems to be exhibited in the loop approach) or vanish as a viable
concept in the final theory. 

In spite of the absence of spacetime, the structure of the
Wheeler--DeWitt equation \eqref{WDW} suggests the introduction of a
novel concept: {\em intrinsic time}, which can be defined by
the local hyperbolic structure of this equation. In contrast to the
Schr\"odinger equation, its kinetic term has the same form as in a
wave equation. The kinetic term thus distinguishes 
a timelike variable by the presence of different signs. One can show
that the timelike sign occurs for the local size (as given by the
square root of the determinant of the three-metric, $\sqrt{h}$); in
cosmological examples, it is usually the volume of the universe that plays the
role of intrinsic time, see below. 

This formal structure of the Wheeler--DeWitt equation with its concept
of an intrinsic time has important consequences for the imposition of
boundary data \cite{Zeh}. For a wave equation one usually specifies
the function and its derivative at hypersurfaces of constant time
(here: intrinsic time). We shall encounter some important consequences
of this fact when discussing quantum cosmology below.

A problem related to the problem of time is the ``Hilbert-space
problem'' \cite{OUP}. The standard (``Schr\"odinger'') inner product in quantum
mechanics is conserved {\em in} time $t$, reflecting the
conservation of probability. But do we need such a product in the
absence of an external time? After all, the concepts of probability
and measurement are not obvious ones in a timeless world. Motivated by the
wave structure of the Wheeler--DeWitt equation, one might instead
consider a ``Klein--Gordon inner product'' because such an inner
product is conserved with respect to (intrinsic) time. However, it
possesses the usual problem of such an inner product, which is the
occurrence of negative
probabilities. This would then perhaps lead to the need of a ``third
quantization'' in which the wave functional itself would become an
operator, similar to the necessary transition from relativistic
quantum mechanics to quantum field theory. This would open a Pandora's
box of possibilities which with the current limited status of
understanding should be avoided. It must be emphasized, however, that
at least at a formal level (not discussing potential anomalies) and in
the one-loop approximation, the various inner products lead to an
equivalent formalism if a certain factor ordering is chosen \cite{b1,b2}.

Most of the work in quantum geometrodynamics thus leaves the question
of the inner product open and focuses on topics which are thought to
be independent of it. This is different, for example, in loop quantum
gravity where a consistent (Schr\"odinger-type) inner product exists
at least at the kinematical level, that is, before the constraints are
imposed. A necessary requirement is, of
course, the recovery of standard quantum field theory with its
standard Hilbert-space structure in an approximate limit. This is met
successfully, see Section~3.

\subsection{Relation to covariant quantum gravity}

Quantum geometrodynamics aims to arrive at a
quantum theory of gravity by a direct quantization of Einstein's
theory of general relativity. There are, however, alternative
methods to achieve this goal.
 The oldest is perturbation theory around a fixed (usually
flat) background. Another approach, which is intrinsically
non-perturbative, is path-integral quantization. Such approaches are
called {\em covariant} because they employ a notion of spacetime
covariance as an important ingredient in the formalism (even if at the
end there is no spacetime). 

The question then arises whether there is any connection between the
canonical and covariant approaches \cite{OUP}. This question also
occurs in standard quantum field theory, but becomes more pressing in
quantum gravity because of the absence of spacetime in the canonical
theory. The connection between both approaches is therefore best
understood in the light of the path integral in which one integrates
over the spacetime metric, in analogy to the integration over the
formal particle paths in quantum mechanics. 

The quantum gravitational path integral is formally given by the
following expression,
\be
\lb{HH83}
Z=\int{\mathcal D}g{\mathcal D}\phi\ \E^{\I S[g,\phi]/\hbar}\ ,
\ee
where the integration over ${\mathcal D}g$ includes an integration
over the three-metric as well as lapse function
$N$ and shift vector $N^a$, and where a matter field denoted by $\phi$
has been taken into account. The non-trivial (and not yet fully
solved) issue is, of course, the precise definition of the measure. 
Other contributions to this volume deal with this question.

At the formal level, one can find from
the demand that $Z$ be independent of $N$ and $N^a$ at the
three-dimensional boundaries the result that the path integral must
satisfy the constraints \eqref{WDW} and \eqref{qdiffeo}
\cite{HH}.
In this sense one can disclose a connection between the covariant
(path integral) and the canonical approaches. Of course, to put these
formal derivations on a rigorous footing is far from trivial. 
Most of the work at the rigorous level has thus focused on the one-loop
approximation of the path integral. The corresponding results have
been derived by Andrei Barvinsky in a series of papers, see
\cite{unitary,Barvinsky,ward} and the references therein. They
describe the state of our knowledge about the connection between the
path-integral and the canonical approach.

\subsection{A brief history of quantum geometrodynamics}

The term {\em quantum geometrodynamics} was already used by John
Wheeler to denote quite generally a quantum version of Einstein's
theory, cf. \cite{Wheeler57}. Here, we shall use this term exclusively
for the canonical version of quantum gravity based on the three-metric
and its canonical momentum. The concept should also not be confused
with the name ``quantum geometry'' which is used synonymously for loop
quantum gravity \cite{Ashtekar}. 

The first traces of the canonical formalism can be found in an early
paper by Felix Klein \cite{Klein}, where he discovered that the first
four Einstein equations are ``Hamiltonian'' and ``momentum density''
equations. A general concept for constraints was put forward by
L\'eon Rosenfeld \cite{Rosenfeld}. He found that the first four Einstein
equations are constraints in this general sense. He also discussed the
issue of the consistency conditions in the quantum theory, that is,
that the commutator between the constraints must close on a
constraint. Following the corresponding discussion by Dirac in
\cite{Dirac50}, this requirement is known as Dirac consistency. 

A general formalism for constrained systems was developed by Dirac in
his papers \cite{Dirac50} and \cite{Dirac58}. In \cite{Dirac58b} he
applied it to the gravitational field and essentially derived, in fact, the
equations \eqref{H} and \eqref{Da}. He also discussed the
reduced-quantization approach. 
Important contributions to canonical gravity came in addition from
Peter Bergmann's group (see e.g. his short review in \cite{Bergmann1}) 
and from Arnowitt, Deser, and Misner (summarized in \cite{ADM}). The
latter gave, in particular, a rigorous definition of gravitational
energy and radiation by canonical methods. As has been mentioned
above, the notion of an observable in this context is due to
Bergmann. Moreover, in 1966 he noted that the wave functional in
canonical quantum gravity (in fact, in general constrained systems of
this kind) is timeless \cite{Bergmann2}. To quote him: ``To this
extent the Heisenberg and Schr\"odinger pictures are indistinguishable
in any theory whose Hamiltonian is a constraint.'' He did not,
however, discuss the explicit form of the quantum constraints
\eqref{WDW} and \eqref{qdiffeo}. 

This was then achieved in the already mentioned papers by John Wheeler
and Bryce DeWitt \cite{battelle,DeWitt}. While the general formalism
was discussed extensively in \cite{DeWitt}, conceptual issues
form the main part of \cite{battelle}. In fact, the Wheeler--DeWitt
equation (as it was, of course, only later called) can be found 
in \cite{battelle} only in an
appendix and in a shorthand notation. However, in his pioneering paper
\cite{DeWitt} DeWitt acknowledges John Wheeler's important influence: ``The
present paper is the direct outcome of conversations with Wheeler,
during which one fundamental question in particular kept recurring:
{\em What is the structure of the domain manifold for the
  quantum-gravitational state functional?}'' (see \cite{DeWitt},
p.~1115). In fact, much space in \cite{DeWitt} is devoted to the
configuration space, the inner product (for which he suggested to use
the Klein--Gordon inner product), but also to the semiclassical limit
and, for the first time, to quantum cosmology. He suggests a first
criterion of singularity avoidance in demanding that the wave function
vanish in the region of a classical singularity.
 DeWitt also addresses the problem of
the interpretation of quantum theory in the light of cosmology, which
motivates him to adopt the Everett interpretation. 

This concludes the early history of quantum geometrodynamics. From
1968 on, the work in this field concentrates on the general issues and
models which are the topic of my contribution. It is somewhat
surprising that Dirac, who contributed so much to the early
development of the field, seems to have lost interest. In a
contribution to a conference which took place in Trieste in 1968 he
gave a talk entitled ``The quantization of the gravitational field''
\cite{Dirac68}. In it he mentions only his own work and a paper by
Schwinger and focuses attention to the open problem of the constraint
algebra, concluding that ``the problem of the quantization of the
gravitational field is thus left in a rather uncertain state''
(\cite{Dirac68}, p.~543). This is perhaps due to his instrumentalist
attitude towards physics (in addition to his emphasis on mathematical
beauty) which forbade him to continue with a physical investigation
before these consistency conditions were solved. Even in such a
small field as geometrodynamics, the tastes of the contributors are
highly diversive. It should, however, be remarked that at least at a
formal level (without addressing the question of regularization), the
factor ordering can be fixed by the requirement that different
quantization approaches be equivalent \cite{b1,b2}.   


\section{The bridge to quantum theory on a fixed background}

\subsection{Hamilton--Jacobi equation}

The fundamental quantum equations \eqref{WDW} and \eqref{qdiffeo}
are usually derived from a three-plus-one decomposition of the
classical spacetime and the imposition of heuristic quantization
rules. One may, however, arrive at those equations from a different
conceptual direction, which is analogous to Schr\"odinger's original
derivation of his famous wave equation. Let us quote Schr\"odinger himself:

\begin{quote}
\ldots {\em we know today, in fact, that our classical mechanics fails for
  very small dimensions of the path and for very great curvatures.}
Perhaps this failure is in strict analogy with the failure of
geometrical optics \ldots that becomes evident as soon as the
obstacles or apertures are no longer great compared with the real,
finite, wavelength. \ldots Then it becomes a question of searching
for an `undulatory mechanics' -- and the most obvious way is by an
elaboration of the Hamiltonian analogy on the lines of undulatory 
optics.\footnote{{\em wir
  wissen doch heute, da\ss\ unsere klassische Mechanik bei sehr kleinen
  Bahn\-dimensionen und sehr starken Bahnkr\"ummungen versagt}. Vielleicht
ist dieses Versagen eine volle Analogie zum Versagen der geometrischen
Optik \ldots, das
bekannt\-lich eintritt, sobald die `Hindernisse' oder `\"Offnungen' nicht
mehr gro\ss\ sind gegen die wirkliche, endliche Wellenl\"ange. 
\ldots Dann gilt es, eine `undulatorische
Mechanik' zu suchen -- und der n\"achstliegende Weg dazu ist wohl die
wellentheoretische Ausgestaltung des Hamiltonschen
Bildes. \cite{Schroedinger}} 
\end{quote}

The essential idea here is to ``guess'' a wave equation that yields
the Hamilton--Jacobi equation of classical mechanics in an appropriate
limit. We can try the same for general relativity: ``guess'' a wave
equation that gives in the classical limit Einstein's equations in
their Hamilton--Jacobi version. But what is the Hamilton--Jacobi
version of these equations? Asher Peres derived it in 1962
\cite{Peres}: instead of the ten Einstein field equations, which are
partial differential equations, one gets the following four
functional differential equations, which are nothing but the four
constraint equations \eqref{H} and \eqref{Da} in the Hamilton--Jacobi form,
\bea
\lb{HJ}
16\pi G\, G_{abcd}\frac{\delta S}{\delta h_{ab}}\frac{\delta S}{\delta
h_{cd}}-\frac{\sqrt{h}}{16\pi G}(\,{}^{(3)}\!R-2\Lambda) &=& 0\ ,\nonumber
\\
D_a\frac{\delta S}{\delta h_{ab}} &=& 0\ .
\eea
(Restriction has here been made to the vacuum case.)
The eikonal $S$ is a functional of the three-metric, $S[h_{ab}(\bf x)]$.
Using the principle of constructive interference, Ulrich Gerlach has
shown in 1969 that the equations \eqref{HJ} are indeed fully
equivalent to all ten Einstein field equations \cite{Gerlach}; this
approach to Einstein's theory is one of the six routes to
geometrodynamics presented in \cite{MTW}.

If one now looks for wave equations for a wave functional
$\Psi[h_{ab}(\bf x)]$ which lead to \eqref{HJ} in the semiclassical
limit, that is, when $\Psi$ is of the WKB form
\be
\Psi[h_{ab}]=C[h_{ab}]\exp\left(\frac{\I}{\hbar}S[h_{ab}]\right)\ ,
\ee
with a slowly varying amplitude $C$ and a rapidly varying phase $S$,
one arrives at the quantum constraint equations \eqref{WDW} and
\eqref{qdiffeo}. 

Independent of their status at the most fundamental level, therefore,
one can argue that the equations \eqref{WDW} and \eqref{qdiffeo}
should at least be valid approximately for energies below the Planck
scale. This conclusion is based only on two rather conservative assumptions: the
universality of the quantum framework (that is, the universal validity
of the superposition principle) and the validity of Einstein's
equation in the classical limit. Both of these assumptions enjoy
strong support: general relativity has passed all experimental and
observational tests so far, and the same is true for quantum theory 
where interference experiments can be extended far into the mesoscopic
regime and where the emergence of classical behaviour is understood as
arising from decoherence \cite{deco,Schlosshauer}.

\subsection{Semiclassical approximation}

The discussion in the last subsection suggests that the semiclassical
limit from quantum geometrodynamics is well understood at least at the
level of the formal constraint equations \eqref{WDW} and
\eqref{qdiffeo}. This is indeed the case \cite{OUP}. One can derive
the limit of quantum field theory in an external spacetime through a
kind of Born--Oppenheimer approximation scheme. This idea was first
spelled out by Lapchinsky and Rubakov \cite{LaRu}.

Starting point is the following ansatz for a general solution of 
\eqref{WDW} and \eqref{qdiffeo}:
\be
\lb{Ansatz}
        \vert\Psi[h_{ab}]\rangle=C[h_{ab}]
        \E^{\I m_{\rm P}^2S[h_{ab}]}|\psi [h_{ab}]\rangle\ ,
\ee
where the bra-ket notation of the wave functional refers to the
standard Hilbert space of non-gravitational degrees of freedom and
where $m_{\rm P}$ is the Planck mass. Inserting this into 
\eqref{WDW} and \eqref{qdiffeo} and performing an expansion with
respect to the Planck mass, one finds in the highest-order
approximations that $S$ obeys \eqref{HJ} and that $\psi
[h_{ab}]\rangle$ obeys 
\bea
\lb{BO}
\left(\hat{\mathcal H}_{\perp}^{\rm m}-\langle\psi\vert\hat{\mathcal H}_{\perp}
^{\rm m}\vert\psi\rangle -\I G_{abcd}\frac{\delta S}{\delta h_{ab}}
\frac{\delta}{\delta h_{cd}}\right)\vert\psi[h_{ab}]\rangle &=&0\ ,
\nonumber \\ 
\left(\hat{\mathcal H}_a^{\rm m}-\langle\psi\vert\hat{\mathcal H}_a^{\rm m}
\vert\psi\rangle-\frac{2}{\I}h_{ab}D_c\frac{\delta}{\delta h_{bc}}
\right)\vert\psi[h_{ab}]\rangle &=&0\ .
\eea
One now evaluates $|\psi[h_{ab}]\rangle$ along a
solution of the classical Einstein equations, $h_{ab}({\mathbf x},t)$,
corresponding to a solution, $S[h_{ab}]$, of the Hamilton--Jacobi
equations \eqref{HJ}; this solution is obtained from
\be
        \dot{h}_{ab}=NG_{abcd}
        \frac{\delta S}{\delta h_{cd}}+
        2D_{(a}{N_{b)}}\ ,
\ee
which is the analogue in relativity of the equation
$\dot{q}=m^{-1}\partial S/\partial q$ in classical mechanics.
Defining a time parameter $t$ by
\bdm
        \frac{\partial}{\partial t}\,|\psi(t)\rangle=
        \int \D^3 x \,\dot{h}_{ab}({\bf x},t)\,
        \frac{\delta}{\delta h_{ab}({\bf x})}
        |\psi[h_{ab}]\rangle\ ,
        \edm
one can derive from \eqref{BO} the following
functional Schr\"odinger equation for the quantized
non-gravitational fields in the chosen external classical gravitational field:
        \begin{eqnarray}
        \lb{Schroedinger}
        \I\hbar\frac{\partial}{\partial t}\,
        |\psi(t)\rangle &=& \hat{H}{}^{\rm m}|\psi(t)\rangle\ ,\nonumber
        \\
        \hat{H}{}^{\rm m} &\equiv&
        \int \D^3 x \left\{N({\bf x})
        \hat{\mathcal H}{}^{\rm m}_{\perp}({\bf x})+
        N^a({\bf x})\hat{\mathcal H}{}^{\rm m}_a({\bf x})\right\} \ ,
               \end{eqnarray}
where $\hat{H}{}^{\rm m}$ is the Hamiltonian for the non-gravitational fields
in the Schr\"o\-ding\-er
picture, which depends parametrically on the (generally non-static) metric
coefficients of the curved spacetime background.
It this level of approximation, the
``WKB time $t$'' controls the dynamics -- time has been regained from
timeless quantum gravity in an appropriate limit.

Together with the parameter $t$, the imaginary unit $\I$ has appeared
in \eqref{Schroedinger}. This entails then the use of the complex wave
functions in quantum theory, which are so essential for its
formalism. But has this not been introduced by hand through the
special ansatz \eqref{Ansatz}? In a certain sense, yes. However, one
can show that superpositions of such complex wave functions become
dynamically independent from each other through decoherence \cite{deco}. 

Consider, for example, a superposition of a state of the form
\eqref{Ansatz} with its complex conjugate. Taking into account
inhomogeneous degrees of freedom such as density fluctuations or weak
gravitational waves, one can show that the resulting entangled state
exhibits only a tiny interference factor between the $\exp(\I S/\hbar)$- and
the $\exp(-\I S/\hbar)$-component of the total quantum state after the
inhomogeneous degrees of freedom have been traced out. 
This is the effect of decoherence. In one example
which I calculated some time ago, the decoherence factor responsible
for this suppression of interference turned out to read \cite{CK92}
\bdm
\exp\left(-\frac{\pi m
    H_0^2a^3}{128\hbar}\right)\sim\exp\left(-10^{43}\right)\ ,
\edm
where $a$ is the scale factor of a Friedmann universe (see below),
$H_0$ the Hubble constant, and $m$ the mass of a scalar field used in
this model. The numerical value arises after some standard values 
for the parameters are
inserted. The smallness of this number means that our present Universe
can be treated as behaving classically to a high degree of accuracy.   

One can interpret this result also as follows. The full quantum
equations \eqref{WDW} and \eqref{qdiffeo} are real equations and
are therefore invariant under complex conjugation. The state
\eqref{Ansatz}, on the other hand, is complex, violating this
symmetry. Since the time parameter $t$ only follows from such a
complex state (which can be interpreted as a decohered branch of a
full real state), one can say that time itself emerges from symmetry
breaking. 

The situation is analogous to molecular physics where the chiral
behaviour of molecules (e.g. sugar molecules) can emerge through a
similar symmetry-breaking effect: while the fundamental equation (the
Schr\"odinger equation with the Hamilton operator for the molecules)
is parity-invariant, the chiral states are not. The dynamical reason
for this symmetry breaking
is again the process of decoherence, there caused by the
scattering with light or air molecules.

\subsection{Quantum gravitational corrections}

If the functional Schr\"odinger equation can be recovered from full
quantum gravity in an appropriate limit, the question arises whether
one can go beyond this limit and calculate quantum gravitational
correction terms. This can be done at least at a formal level, that
is, at the level where one treats the functional derivatives like
partial derivatives. 

The next order in the Born--Oppenheimer approximation then gives
corrections to the Hamiltonian for the non-gravitational fields,
\be
\hat{H}{}^{\rm m} \to
\hat{H}{}^{\rm m} +\frac{1}{m_{\rm P}^2}\left({\rm various\
    terms}\right)\ .
\ee
The detailed form of these terms can be found in
\cite{OUP,KS,BK}. Future investigations should deal with a concrete
application of these terms in cosmology, for example, in the search
for quantum gravitational effects in the anisotropy spectrum of the
Cosmic Background Radiation.

A simple example is the calculation of the quantum gravitational
correction to the trace anomaly in de~Sitter space \cite{CK96}.
For a conformally coupled scalar field, the trace of the
energy--momentum tensor, although being zero classically, is
non-vanishing in the quantum theory; this ``anomalous trace'' is
proportional to $\hbar$. It corresponds to the following expectation
value, $\varepsilon$, of the Hamiltonian density, 
\be
\varepsilon=\frac{\hbar H_{\rm dS}^4}{1440\pi^2c^3}\ ,
\ee
where $H_{\rm dS}$ is the constant Hubble parameter of de~Sitter
space. The first quantum gravitational correction calculated from the
Born--Oppenheimer expansion discussed above reads
\be
\delta\varepsilon\approx -\frac{2G\hbar^2H_{\rm
    dS}^6}{3(1440)^2\pi^3c^8}\ ,
\ee
so that the ratio is given by 
\be 
\lb{ratio}
\frac{\delta\varepsilon}{\epsilon}\approx
-\frac{1}{2160\pi}\left(\frac{t_{\rm P}}{H_{\rm dS}^{-1}}\right)^2\ ,
\ee
where $t_{\rm P}$ denotes the Planck time. One might perhaps have
guessed for dimensional reasons that the ratio of the Planck time to
the Hubble time enters, but this example shows that in principle exact
results can be obtained from canonical quantum gravity. Numerically,
the ratio \eqref{ratio} is, of course, small. Using values motivated
by inflationary cosmology, one can assume that $H_{\rm dS}$ lies
between $10^{13}$ and $10^{15}$ GeV, leading for the ratio
\eqref{ratio} to values between roughly $10^{-16}$ and $10^{-22}$. 
It is at present an open question whether there are relevant cases
where the correction terms can be big enough to be observable.


\section{Quantum black holes and quantum cosmology}

\subsection{Quantum black holes}

According to the no-hair theorem of general relativity, stationary
black holes are uniquely characterized by the three parameters mass,
angular momentum, and electric charge. If all parameters are
non-vanishing, the solution is given by the Kerr--Newman metric, which
is axially symmetric. Most investigations into the quantum aspects
have focused on the simple situation of vanishing angular momentum,
because then the solutions are spherically symmetric. Still, the
difficulties in performing the quantization are formidable. 

The simplest case is the eternal Schwarzschild black hole without
matter degrees of freedom. Such a black hole is fully characterized by
its mass, $M$.
Through a series of sophisticated
transformations, Karel Kucha\v{r} was able to reduce the problem to a
purely quantum mechanical one and give an explicit form of the
resulting wave function \cite{Kuchar2}. 
If one extends this solution to include an electric charge $q$, the wave
function reads (see e.g. \cite{LWH,OUP})
\begin{equation}\label{psi-solution}
  \Psi (\alpha, \tau, \lambda) = \chi (M,q) \exp \left[
  \frac{\I}{\hbar} \left( \frac{A(M,q) \alpha}{8 \pi G}- M\tau -q\lambda
  \right) \right]\ ,
\end{equation}
where $\chi(M,q)$ is an arbitrary function of $M$ and $q$, $A(M,q)$ is
the area of the horizon as expressed through mass and charge, 
$\lambda$ is a parameter conjugated to charge, $\alpha$ a `rapidity
parameter' connected with the bifurcation sphere of the black-hole
horizons in the Kruskal diagramme, and
$\tau$ denotes the Schwarzschild (Killing) time at asymptotic
infinity. In contrast to the general case discussed above, such a time
variable is available in the asymptotic regime of an asymptotically
flat situation, that is, far away from the black hole. If additional
matter is present, such a reduction to finitely many degrees of
freedom is no longer possible and one has to deal with the full
functional equations.

It is possible to discuss a quantum state for the black hole in a
one-loop approximation. Choosing such a state in accordance with the
no-boundary state in quantum cosmology (see below), Barvinsky {\em et
  al.} have calculated the entanglement entropy arising from this
state when all the degrees of freedom outside the horizon are traced
out \cite{BFZ}. They found for the entropy the expression
\be
S=-k_{\rm B}{\rm Tr}(\rho\ln\rho)=k_{\rm B}\frac{A}{360\pi l^2}\ ,
\ee
where $\rho$ is the density matrix resulting from tracing out the
exterior degrees of freedom, and $l$ is a cutoff parameter denoting
the proper distance to the horizon. One recognizes that this
expression is divergent for $l\to 0$. This calculation is therefore
not yet a complete one; on the other hand, it yields the expected
proportionality between black-hole entropy and area. 

In the attempt to recover the Bekenstein--Hawking entropy $S_{\rm BH}$ from an
entanglement entropy, one has to keep in mind the universality of 
$S_{\rm BH}$, that is, its independence from the actual field
content. What could give such a universality? One universal feature
of a black hole is the spectrum of its quasi-normal modes, which are
damped out when reaching the stationary black-hole state, but which
could still play a role in the quantum theory. They stay entangled
with the black hole and tracing them out could perhaps give $S_{\rm
  BH}$ \cite{Kiefer04}. However, any serious calculation is elusive. 

Instead of an eternal black hole one can attempt to describe a black
hole that results dynamically from a gravitational collapse. One
example is a collapsing spherically symmetric dust shell. Classically,
it collapses to form a black hole. In the quantum theory, interesting
features can happen \cite{Hajicek}. If the shell is described by a
narrow wave packet, it turns out that this packet will first collapse,
enter slightly inside the classical event horizon and then re-expand
to infinity. In a sense, the quantum theory yields a superposition of
a black-hole with a white-hole solution, resulting in a destructive
interference of the total wave packet in the region of the classical
singularity: for $r\to0$, the wave function obeys $\Psi\to 0$. This
is a consequence of constructing a unitary (with respect to asymptotic
time) canonical quantum theory. 

Instead of a dust shell, one can consider a spherically symmetric dust
cloud -- the Lema\^{\i}tre--Tolman--Bondi (LTB) model. Classically,
this is a self-gravitating dust cloud with energy--momentum tensor 
$T_{\mu \nu} = \epsilon(\tau,\rho) u_{\mu} u_{\nu}$ and is given by
the line element 
\be
\mathrm{d}s^2 = -\mathrm{d}\tau^2 +
\frac{(\partial_{\rho}R)^2}{1+2E(\rho)} \mathrm{d}\rho^2
+ R^2(\rho)(\mathrm{d}\theta^2 + \sin^2\theta \mathrm{d}\phi^2)\ .
\ee
The canonical formalism and its quantization were developed by Vaz
{\em et al.} in \cite{VWS}. After some manipulations both the
Wheeler--DeWitt equation and the diffeomorphism constraint (in the
case of spherical symmetry there is only one such constraint) were
presented in a simplified, but still functional, form.
 
In a series of paper, the following results were obtained (see
\cite{VGKSW} and the references therein). Firstly, exact quantum
states of a particular type were found. This is possible because the
dust shell can be imagined as being composed of infinitely many
decoupled shells. The exact quantum states, which can be found only in
a special factor ordering, can be interpreted as an infinite product of
single-shell states. Although being exact solutions, they are of a WKB
form. Secondly, it was possible to retrieve from these quantum
gravitational states the standard expressions for the Hawking
radiation {\em plus} explicit corrections due to greybody factors.  
For the BTZ black hole, which is a solution in 2+1 dimensions with
negative cosmological constant $\Lambda$, 
it was possible to derive the
Hawking temperature and to give a microscopic derivation of the black-hole
entropy. In fact, it was found in this 2+1-dimensional case that there
is a discrete mass spectrum for the shells collapsing to the black
hole. 

Following early suggestions by Jacob Bekenstein,\footnote{``It is
 then natural to 
introduce the concept of black-hole entropy as the measure of the {\it
  inaccessibility}  
of information (to an exterior observer) as to which particular
internal configuration  
of the black hole is actually realized in a given case.'' \cite{Bekenstein}}
the black-hole
entropy is there defined as the number of possible distributions of
${\mathcal N}$ identical shells between these levels. The result is
 \be
S_\text{can}\approx 2\pi k_{\rm B}
\sqrt{\left(1-\frac{48lM_0}{\hbar}\right)\frac{lM}{6\hbar}}\ ,
\ee
where $l\equiv\vert\Lambda\vert^{-1/2}$, $M$ is the mass of the BTZ black
hole, and $M_0$ is a free constant of the model. This entropy is equal
to the Bekenstein--Hawking entropy if this constant is chosen as follows:
\be
M_0 = - \frac 1{16G} + \frac{\hbar}{48l}\ .
\ee
Actually, $M_0$ can be related with the conformal charge of the
effective con\-for\-mal-field theory usually used to derive the entropy
for the BTZ black hole, cf. \cite{Carlip}. All of these results are,
of course, preliminary, but they demonstrate to which extent quantum
geometrodynamics can be applied in the understanding of black holes.

\subsection{Quantum cosmology}

Quantum cosmology is one of the main applications of quantum
geometrodynamics. Its purpose is twofold: On the one hand, it can
serve as a toy model for full quantum gravity in which the
mathematical difficulties disappear. On the other hand, it can be
employed as a description for the real Universe, with the final goal
to be tested by observation. 

In this subsection, I shall focus on some recent work into which I was
myself involved. More detailed overviews of quantum cosmology can be
found, for example, in \cite{OUP,Halliwell,Wiltshire,Coule,KS}. 

The simplest model of quantum cosmology is the quantization of a
Fried\-mann--Lema\^{\i}tre universe. The classical line element is taken
to be of the form
\be
\D s^2=-N^2(t)\D t^2+a^2(t)\D\Omega_3^2\ ,
\ee
where $N$ is the lapse function, $a$ the scale factor, and we have
chosen the three-dimensional space to be closed. In addition, we shall
implement a homogeneous matter field $\phi$ as a representative for
matter. We are thus left with a two-dimensional configuration space
(consisting of $a$ and $\phi$); because of the huge truncation of
the infinite-dimensional superspace, such a space is called
minisuperspace. 

The diffeomorphism
constraints are identically satisfied by this ansatz, and 
the Wheeler--DeWitt equation reads (with units $2G/3\pi=1$ and $c=1$)
\be
\lb{mini}
\frac{1}{2}\left(\frac{\hbar^2}{a^2}\frac{\partial}{\partial a}
\left(a\frac{\partial}{\partial a}\right)-\frac{\hbar^2}{a^3}
\frac{\partial^2}{\partial\phi^2}-a+\frac{\Lambda a^3}{3}+
m^2a^3\phi^2\right)\psi(a,\phi)=0\ . 
\ee
The factor ordering has been chosen to be of the Laplace--Beltrami
form, which has the advantage that it guarantees covariance in
minisuperspace. 

It is evident that equations such as \eqref{mini} do not possess the
mathematical problems of the full functional equation \eqref{WDW}.
One can thus focus attention on physical applications. One important
application is the imposition of boundary conditions. Popular
proposals are the no-boundary condition \cite{HH} and the
tunneling condition \cite{Vilenkin}. The no-boundary proposal makes
essential use of the connection between covariant and canonical
quantum gravity discussed in Section~2.6: it is defined conceptually
by a Euclidean path integral, but also relies on solving a
minisuperspace Wheeler--DeWitt equation such as \eqref{mini}. Other
important applications include the discussion of wave packets, the
validity of the semiclassical approximation, the origin of classical
behaviour and the arrow of time, and the possible quantum avoidance of
classical singularities \cite{OUP,Zeh}.  

Before picking out one particular model, I want to emphasize one
important conceptual point which is relevant for the problem of time
discussed above, see Figure~2. 

\begin{figure}[h]
\begin{center}
\includegraphics[width=5cm]{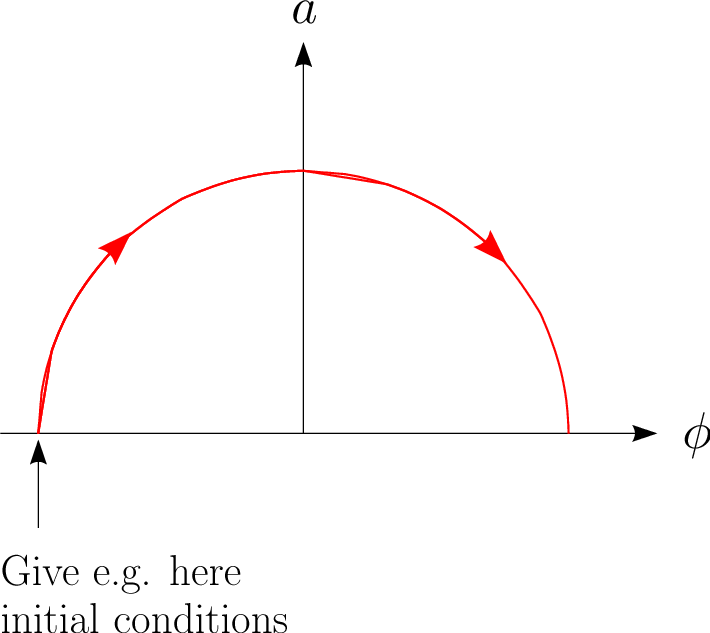}
\includegraphics[width=5cm]{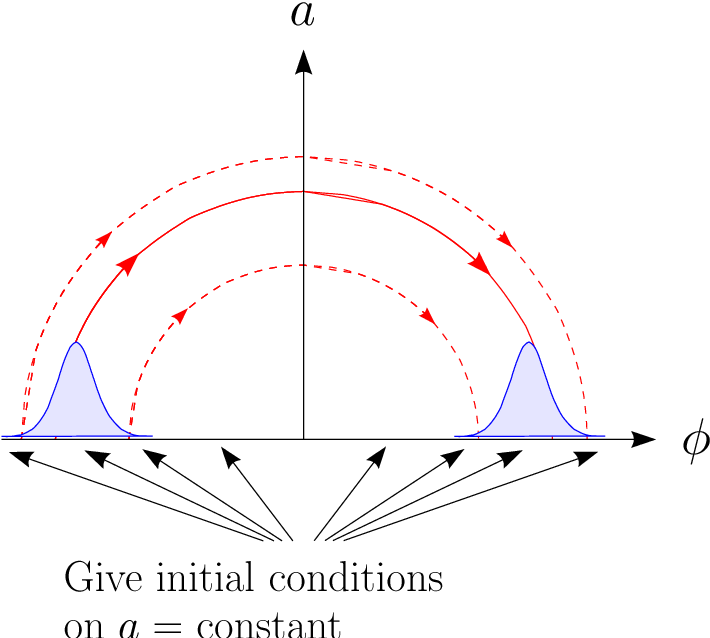}
\caption{The classical and the quantum theory of gravity exhibit
drastically different notions of determinism \cite{OUP}.}
\end{center}
\end{figure}

Consider a two-dimensional
minisuperspace model with the variables $a$ and $\phi$ as above. The
figure on the left shows the classical trajectory in configuration
space for a universe which is expanding and recollapsing. Classically,
one can give initial conditions, for example, on the left end of the
trajectory for small $a$ and then determine the whole trajectory. In
this sense, the recollapsing part of the trajectory is the
deterministic successor of the expanding part. One could, of course,
also start from the right end of the trajectory because there is no
distinguished direction; but the important point is that a trajectory
exists. Not so in the quantum theory where both the trajectory and the
time parameter $t$ are absent! If one wants to find a solution of the
Wheeler--DeWitt equation which describes a wave packet following the
classical trajectory, one has to specify two packets at the would-be
ends of the classical trajectory, see the right figure. The reason is
that \eqref{mini} is a hyperbolic equation with respect to intrinsic
time $a$, and the natural formulation of boundary conditions is to
impose the wave function (and its derivative) at constant $a$. If one
imposed only one of the two wave packets, the full solution would be a
smeared-out wave function which does not resemble anything like a wave packet
following the classical trajectory. In this sense, the
``recollapsing'' wave packet must be present ``initially''. 

Quantum geometrodynamics thus provides us with crucial insights into
the nature of time in quantum gravity. And the consequences of this
new concept of time are independent of 
any particular scale, that is, independent of possible
modifications of the theory at the Planck scale. 

Let us now turn to a
specific example \cite{KKS}: a cosmological model with a ``big
brake''. Classically, the model is characterized by an 
equation of state of the form $p=A/\rho$, where $A>0$ (``
anti-Chaplygin gas''). This can be realized by a scalar field $\phi$ with the
following potential (with $\kappa^2=8\pi G$):
\be
V(\phi)=V_0\left(\sinh{\left(\sqrt{3\kappa^2}|\phi|\right)}-
\frac1{\sinh{\left(\sqrt{3\kappa^2}|\phi|\right)}}\right)\ ; \ V_0=\sqrt{A/4}\ .
\ee
This model universe develops a pressure singularity at the end of its
evolution where it comes to an abrupt halt: $\dot{a}$ remains finite
there but $\ddot{a}(t)$ tends to minus infinity; this is why it is called a
``big brake''. Since this model does not describe an accelerating
universe, it is as such in conflict with present
observations. However, it can easily be generalized in order to
accommodate such an acceleration, without modifying the following discussion.
The total lifetime of this universe is
\bdm
t_0\approx 7\times 10^{2}\frac{1}{\sqrt{V_0\left[\frac{\rm g}{{\rm
          cm}^3}\right]}}\ {\rm s}\ ,
\edm
which is much bigger than the current age of our Universe for
\bdm
V_0\ll 2.6\times 10^{-30}\ \frac{\rm g}{{\rm cm}^3}\ .
\edm
The classical trajectory in configuration space is shown in Figure~3. 
The big-brake singularity is at $\phi=0$. In addition, there are the
usual big-bang and big-crunch singularities at $a=0$ and
$\phi\to\pm\infty$.
\begin{figure}[h]
  \begin{center} 
  \includegraphics[width=0.7\textwidth]{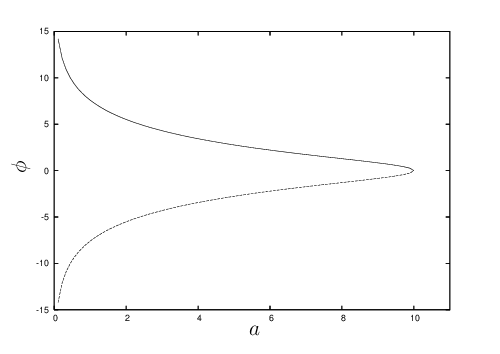} 
  \caption{Classical trajectory in configuration space \cite{KKS}.}
  \end{center}
\end{figure}

In the quantum theory, one encounters the following Wheeler--DeWitt
equation: 
\bea
& &\frac{\hbar^2}{2}\left(\frac{\kappa^2}{6}\frac{\partial^2}
{\partial\alpha^2}-\frac{\partial^2}{\partial\phi^2}\right)\Psi\left(\alpha,
  \phi\right)\nonumber \\
&+&V_0e^{6\alpha}\left(\sinh{\left(\sqrt{3\kappa^2}|\phi|\right)}-
\frac1{\sinh{\left(\sqrt{3\kappa^2}|\phi|\right)}}\right)\Psi\left(\alpha, 
  \phi\right)=0 \ ,
\eea
 where $\alpha=\ln a$, and a Laplace--Beltrami
 factor ordering has again been employed. In order to study the behaviour
 near the region of the classical singularity, it is sufficient to
 study the limit of small $\phi$. One can then use the approximate
 equation 
\be
\frac{\hbar^2}2\left(\frac{\kappa^2}{6}\frac{\partial^2}{\partial\alpha^2}
-\frac{\partial^2}{\partial\phi^2}\right)\Psi\left(\alpha,
  \phi\right)-\frac{\tilde{V_0}}{|\phi|}e^{6\alpha}\Psi\left(\alpha,
  \phi\right)=0\ ,
\ee
where $\tilde{V_0}={V_0}/{3\kappa^2}$. A crucial input is now the
choice of boundary conditions. Firstly, we have to impose the condition
that the wave function go to zero for large $a$; this is because the
classical evolution stops at finite $a$. Secondly, we demand
normalizability with respect to $\phi$. The resulting solutions are
then of the form
\be
\Psi\left(\alpha,\phi\right)=
\sum_{k=1}^{\infty}A(k)k^{-3/2}\mathrm{K}_0
\left(\frac{1}{\sqrt{6}}\frac{V_{\alpha}}{\hbar^2k\kappa}\right)
\left(2\frac{V_\alpha}{k}|\phi|\right)
e^{-\frac{V_\alpha}{k|\phi|}}\mathrm{L}^1_{k-1}
\left(2\frac{V_\alpha}{k}|\phi|\right)\ ,
\ee
where $\mathrm{K}_0$ is a Bessel function, $\mathrm{L}^1_{k-1}$
denotes the Laguerre polynoms, and
$V_\alpha\equiv\tilde{V_0}e^{6\alpha}$. Inspection of this solution
shows that it {\em vanishes} at $\phi=0$, that is, at the classical
big-brake singularity. Therefore, 
this singularity is avoided in the quantum theory. 
In fact, the normalization condition with respect to $\phi$ also
guarantees that the big-bang singularity is absent. One is thus left
with a singularity-free quantum universe. 

A wave-packet solution following the classical solution of Figure~3 and
approaching zero when $\phi\to 0$ (that is, when approaching the
region of the classical big-brake singularity), is shown in Figure~4. 
\begin{figure}[h]
  \begin{center} 
  \includegraphics[width=0.8\textwidth]{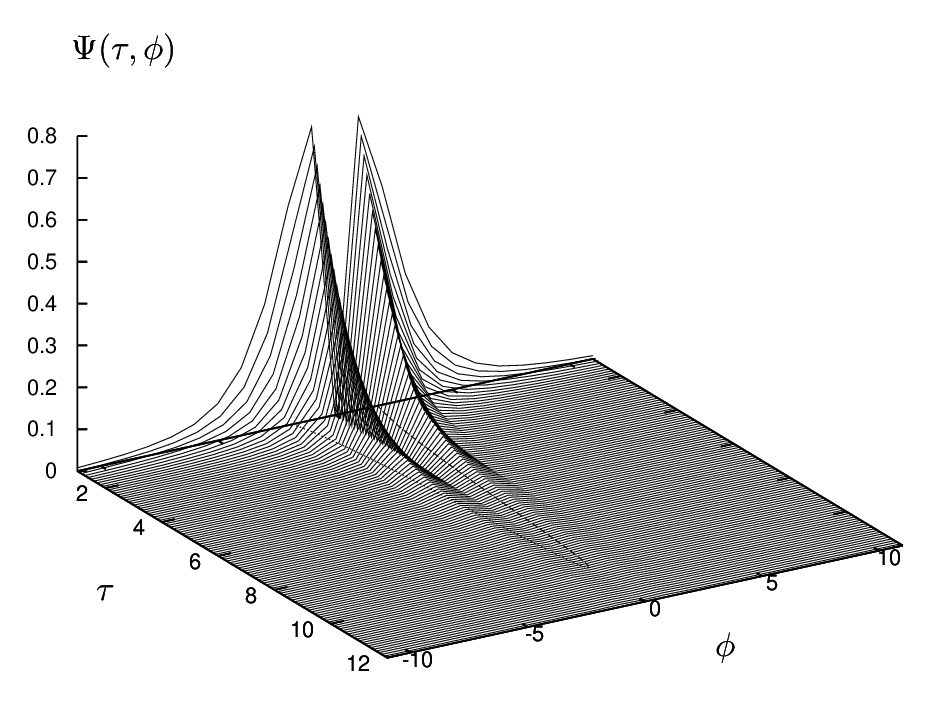} 
  \caption{The wave packet for the big-brake model. The packet follows
  the classical trajectory but becomes zero at the classical
  singularity \cite{KKS}.}
  \end{center}
\end{figure}

A somewhat related model with a quantum avoidance is phantom cosmology
\cite{DKS}. Classically, one has there a universe with scale factor $a(t)$
containing a scalar field with negative kinetic term (``phantom''), which
develops a ``big-rip singularity'':
$\rho$ and $p$ diverge as $a$ goes to infinity at a {\em finite time}.
An investigation of the Wheeler--DeWitt equation demonstrates that
wave-packet solutions 
{\em disperse} in the region of the classical big-rip singularity.
Therefore, time and the classical evolution come to an end before the
singularity would be reached. 
Only a stationary quantum state is left. This, again, presents an
example where quantum gravitational effects are important for large
scale factor -- much bigger than the Planck length. Quantum geometrodynamics is
able to cope with this situation.  

Quantum cosmology extends well beyond the minisuperspace limit of
homogeneity \cite{OUP}. In order to understand structure formation, it
is crucial to implement inhomogeneous perturbations \cite{HaHa}. The
tensor part of these perturbations then describes weak quantized
gravitational waves. It is also of interest to investigate a quantum
analogue of the Belinski--Khalatnikov--Lifshitz analysis of approaching
a spacelike singularity. It has been argued that this leads, in
addition to the disappearance of time, to an effective de-emergence of
space \cite{DN}. The classical singularity would then be fully dissolved
in quantum gravity. 

All models of quantum cosmology discussed so far are based on the
assumption that the total quantum state (the ``wave function of the
universe'') is a pure state. Recently the idea arose to start instead with a
fundamental density matrix of a microcanonical ensemble 
\cite{landscape}. If defined by a Euclidean path integral, it was
found that such a state is dynamically preferred compared to the
``no-boundary state'' of \cite{HH}. An interesting result of this
investigation is that the cosmological constant would be limited to a
bounded range. 

Quantum geometrodynamics can also be successfully applied to
lower-dimensional gravity. In $2+1$ dimensions, the gravitational
theory is of a purely topological nature and one thus only has to deal with
finitely many degrees of freedom, similar to quantum cosmology
\cite{Carlip2}. One thereby gets important insights in both the role
of boundary conditions and the structure of the Wheeler--DeWitt
equation.


\section{Conclusions and Outlook}

``There is no experimental evidence for the quantization of the
gravitational field, but we believe quantization should apply to all
the fields of physics. They all interact with each other, and it is
difficult to see how some could be quantized and others not.'' This
is, in Dirac's words (\cite{Dirac68}, p.~539), the main motivation for
dealing with quantum gravity. Because there is no experimental
evidence so far, it is not surprising that several different
approaches are being seriously discussed. In my contribution, I have
addressed one of them, quantum geometrodynamics, which is a direct
quantization of Einstein's theory by canonical means and choosing the
three-metric as its canonical configuration variable. As I have tried
to argue, quantum geometrodynamics is still a viable field because
it gives intuitive insights into many conceptual and technical
questions and because it is able to address 
quantum aspects of black
holes and cosmology. And independent of its status as a fundamental
theory (which it is probably not) it should be valid at least
approximately for length scales bigger than the Planck length -- just
because it can 
be constructed from the condition that it give the correct
semiclassical limit. 

The final decision about quantum gravity will, of course, be made by
experiment. Before that state will be reached, it is important to be
open minded and to investigate as many approaches as possible and to
study both mathematical and conceptual aspects. I would like to close
with a remark by Einstein, who emphasized the non-trivial nature of
the relation between theory and experience in clear words:
\begin{quote}
The concepts and sentences only get ``sense'' and ``content'' through
their relation with the sensual experiences. The connection of the
latter with the former is purely intuitive, not itself of logical
nature. The degree of certainty, with which this relation
resp. intuitive connection can be undertaken, and nothing else,
distinguishes the queer illusion from the scientific 
``truth''.\footnote{Die Begriffe und S\"atze erhalten ``Sinn''
  bzw. ``Inhalt'' nur durch 
ihre Beziehung zu den Sinnenerlebnissen. Die Verbindung der letzteren
mit den ersteren ist rein intuitiv, nicht selbst von logischer
Natur. Der Grad der Sicherheit, mit der diese Beziehung bzw. intuitive
Verkn\"upfung vorgenommen werden kann, und nichts anderes, unterscheidet
die leere Phantasterei von der wissenschaftlichen
``Wahrheit''. \cite{Einstein} } 
\end{quote}

\begin{acknowledgements}
I thank the organizers of the 405th Heraeus seminar: {\em Quantum
  Gravity: Challenges and Perspectives} for inviting me to such an
interesting and stimulating meeting. I am grateful to Andrei
Barvinsky, Alexander Kamenshchik, and Barbara Sandh\"ofer for their
comments on my manuscript.   
\end{acknowledgements}



\end{document}